\documentclass[nohyper,notoc]{JHEP3}

\usepackage{amsmath,euscript,array,amssymb,cite,bm} 
\def\tr{{\rm tr}}

\def\sst{\scriptscriptstyle}
\def\det{{\rm det}}

\def\Dbarslash{\,\,{\raise.15ex\hbox{/}\mkern-12mu {\bar\D}}}
\def\Dslash{\,\,{\raise.15ex\hbox{/}\mkern-12mu \D}}
\def\delslash{\,\,{\raise.15ex\hbox{/}\mkern-9mu \partial}}
\def\delbarslash{\,\,{\raise.15ex\hbox{/}\mkern-9mu {\bar\partial}}}

\def\D{{\cal D}}
\def\Dbarslash{\,\,{\raise.15ex\hbox{/}\mkern-12mu {\bar\D}}}
\def\delslash{\,\,{\raise.15ex\hbox{/}\mkern-9mu \partial}}
\def\Dslash{\,\,{\raise.15ex\hbox{/}\mkern-12mu \D}}

\def\={\, =\, }
\def\+{\, +\, }
\def\-{\, -\, }

\newcommand{\be}{\begin{equation}}
\newcommand{\ee}{\end{equation}}
\def\bea{\begin{eqnarray}}
\def\eea{\end{eqnarray}}

\def\Nf{{N_{f}}}
\def\Nc{{N_{c}}}
\def\p{\varphi}
\def\pt{\tilde\varphi}
\def\uno{\mbox{1 \kern-.59em {\rm l}}}

\title{Naturalised Supersymmetric Grand Unification}

\author{Steven A.~Abel, Joerg Jaeckel and Valentin V.~Khoze\\
Institute for Particle Physics Phenomenology and Centre for Particle Theory, \\
University of Durham,
Durham, DH1 3LE, UK\\
{\tt s.a.abel@durham.ac.uk,}\,  
{\tt joerg.jaeckel@durham.ac.uk,}\, 
{\tt valya.khoze@durham.ac.uk}}

\abstract{We construct a simple model of an $SU(5)$ GUT with gauge mediated supersymmetry breaking
from a metastable vacuum of a hidden sector. All mass parameters and hierarchies of our model
are generated dynamically from retrofitting. This includes the $\mu$-parameter and the GUT scale.
However, as typical for simple $SU(5)$ GUT models, proton longevity remains a problem.}
 
\preprint{{\tt hep-ph/0703086}
\\  IPPP/07/05\\
DCPT/07/10
}

\begin{document}

\section{Introduction}

The mechanism whereby supersymmetry is broken in Nature is once again
the subject of intense scrutiny. Of particular importance has been
the realization by Intriligator, Seiberg and Shih (ISS) that (for
appropriate choices of flavours and colours) the simplest SQCD models
have SUSY breaking metastable minima \cite{ISS}.
Such models are phenomenologically acceptable provided the decay time
from the metastable to the supersymmetric vacuum is sufficiently long. 
Furthermore, it was argued that
the early Universe is naturally driven to such metastable minima and remains
trapped there \cite{ACJK,heat2,heat3,heat4,heat5}. Metastability allows for the
presence of supersymmetric `true'  vacua in the theory and thereby
allows one to evade several stringent constraints on supersymmetry breaking.
These  include
the Nelson--Seiberg theorem \cite{Nelson:1993nf}
which requires an $R$-symmetry leading to unwelcome phenomenological
consequences, such as vanishing gaugino masses or the presence of $R-$axions.

Accepting metastable SUSY breaking minima \cite{ISS} (for earlier work 
see \cite{ELR,dimop,Luty})
leads to a far broader class of SUSY breaking models which is very appealing. 
This fact is exploited in the \emph{retrofitting}
approach of \cite{Dine:2006gm,DM} which -- in the light
of the ISS model -- generalizes and greatly improves upon earlier models of
metastable SUSY breaking. The approach begins with
a model that has an exact $R$-symmetry, and breaks it with terms
that are generated dynamically and are thus small. The models are
metastable, but the fact that $R$-symmetry is still approximately
conserved is enough to ensure that the global SUSY preserving minima
are far away in field space and hence the SUSY breaking minima are
long lived. There have since been a number of discussions of how such
metastable SUSY breaking might be mediated to the Standard Model,
including direct mediation \cite{ISS,Kitano:2006xg,Csaki:2006wi},
breaking within the visible sector \cite{Abel:2007uq} and gauge mediation
\cite{Dine:2006gm,DM,MN,AS,Murayama:2007fe}. 

Our purpose in this paper is to examine the consequences of these
developments for Grand Unification. In particular, the retrofitting
programme seeks to explain all mass scales dynamically by the confinement
of hidden gauge sectors. As well as the SUSY breaking scale itself,
one would naturally like to obtain explanations for other dimensionful
parameters such as the $\mu$-term of the MSSM (as in e.g. \cite{DM}).
Grand Unified Theories (GUTs) are of course full of dimensionful parameters:
the GUT scale; the SUSY breaking scale; the $\mu$-term of the effective
low energy theory. In the simplest $SU(5)$ GUTs, the latter is especially
bothersome, requiring a fine-tuning between mass parameters to one
part in $10^{14}$, the so-called doublet-triplet mass-splitting problem
(for a review see \cite{Nilles:1995ci}). One is led to ask whether
GUTs can be made more natural in the light of metastability: is it
possible to retrofit a GUT model with broken supersymmetry entirely,
so that no dimensionful parameters have to be chosen by hand? 

Basing our analysis on the simplest examples of gauge mediation developed
in refs.\cite{MN,AS} we will argue that it is.
The former paper outlined a simple model of gauge mediation,
whereas the latter showed how it can be retrofitted, with all mass
terms being generated dynamically. However, neither considered the
coupling to, or parameters of the MSSM, such as for example the $\mu$-parameter.
Our objective in the present work will be to completely retrofit this
parameter as well as the other parameters required for GUT and SUSY
breaking itself: in other words to construct a theory whose GUT breaking,
SUSY breaking, messenger scale and $\mu$-term are all generated by
the dynamics. 
We will be able to generate and explain within this approach the three key scales of the visible
sector: the GUT scale $\sim 10^{16} \div 10^{17}$ GeV, the
electro-weak and the supersymmetry breaking scales, both $\sim 10^2$ GeV. In particular, 
our model predicts a relation between the GUT and the elctroweak scale,
\be
M^{2}_{\sst GUT}\, \sim\, {4\pi}\left(\mu_{\sst MSSM} M^{3}_{p}\right)^{1/2} \ .
\label{gutmurel-in}
\ee

For this preliminary study we will be considering the simplest case which is 
a minimal $SU(5)$ GUT (for a review see \cite{Raby:2006sk}). 
These models are known to conflict with bounds on 
the decay rate of the proton because of large dimension-5 operators mediated by 
Higgs triplets. Indeed in the model we present here, the Higgs triplets are 
lighter than usual (although still relatively close to the GUT scale) so that the proton 
decay rate is significantly worse. Nevertheless the model is an encouraging 
first step on the road to a fully consistent retrofitted GUT. We discuss in a 
later subsection how the model or similar GUT models may be developed in order
to make it more realistic.


\section{The model} \label{sec:model}

We want to construct a simple and predictive model which combines and inter-relates the
ideas of supersymmetric Grand Unification \cite{SGUT}, 
supersymmetry breaking by a metastable vacuum \cite{ISS},
and naturalness achieved through retrofitting \cite{DM}.

Following the general set-up of \cite{AS} we consider a model made up of three sectors.
\begin{enumerate}
\item The first is the {\it R-sector} whose main r{\^o}le is to dynamically generate all mass-parameters
in the effective Lagrangian of the full model. This is achieved via a version of the retrofitting 
approach of \cite{Dine:2006gm,DM} which will be reviewed shortly. In our model this sector is described by 
a strongly coupled confining SQCD theory with the dynamical scale $\Lambda_{\sst R}$. In the full theory
$\Lambda_{\sst R}$ triggers the dynamical generation of masses as in \cite{AS}. In addition, in our model the 
$N_f \times N_f$ meson superfield $\tilde{Q}_{\sst R} Q_{\sst R}$ 
of the R-sector will play the r{\^o}le of the adjoint
Higgs of the GUT sector.
\item The second sector is responsible for supersymmetry breaking.
It is described by the SQCD in a free magnetic phase, known as the ISS model \cite{ISS}.
This model contains a long-lived metastable vacuum
which breaks supersymmetry, and will be referred to as the metastable susy-breaking, or
{\it MSB-sector}.
\item The visible sector is the $SU(5)$ susy {\it GUT-sector}. The $SU(5)$ gauge group arises from
gauging the flavour $SU(N_f=5)$ symmetry of the R-sector, and the adjoint Higgs field 
$\Phi_{\sst GUT}$ is identified with the traceless part of the R-sector mesons 
$\tilde{Q}_{\sst R} Q_{\sst R}.$
The GUT-sector is coupled to the MSB-sector
via messenger fields $f$ and $\tilde{f}$ which are in the fundamental and the anti-fundamental of the
$SU(5)$ gauge group. Hence supersymmetry breaking is mediated to the GUT theory via gauge mediation.
\end{enumerate}

In what follows we will see that
this model delivers a supersymmetric Grand Unified Theory with calculable soft susy-breaking
terms (arising from interactions with the MSB-sector). 
The model is fully natural and all the mass-scales of the theory are generated in terms
of appropriate combinations of the two dynamical scales $\Lambda_{\sst R}$, $\Lambda_{\sst MSB}$
and the Planck scale $M_p$. In particular, 
by choosing $\Lambda_{\sst R}$ and $\Lambda_{\sst MSB}$
our model can naturally generate the desired values of the
electro-weak, supersymmetry breaking, and the GUT scale.

\subsection{Interactions between the sectors} \label{sec:inter}

Now we proceed to specify the interactions \emph{between} the three sectors of the model.
These are introduced through the superpotentials ${\cal W}_1$, ${\cal W}_2$ and ${\cal W}_3$
with one property in common: they couple bilinears from one sector to a bilinear from another
and as such are represented by lowest-dimensional non-renormalizable operators suppressed by $M_p$.
For simplicity of presentation, in equations \eqref{supone}, \eqref{suptwo}, \eqref{supthree}
we will include only the interactions which are necessary for our model. Other interactions will be
discussed in the Appendix.
The superpotential ${\cal W}_1$ is responsible for the retrofitting \cite{DM,AS} and couples the singlet bilinear made
of the gauge-strength superfield $W_R$ of the confining R-sector to the singlet bilinears of the
MSB- and the GUT-sectors:
\be
{\cal W}_1 \, =\,
 \tr(W_R^2) \left(
\frac{1}{g_{\sst R}^2}\, +\, \frac{a_{1}}{16 \pi^2 M_p^2}\,\tr(\tilde{Q}_{\sst MSB} Q_{\sst MSB})
\,+ \,\frac{a_{2}}{16 \pi^2 M_p^2}\,\tr(\tilde{f}f) \,
+ \,\frac{a_{3}}{16 \pi^2 M_p^2}\,\tr(\tilde{H}H)\right) \ ,
\label{supone}
\ee
where $\tilde{Q}_{\sst MSB}$, $Q_{\sst MSB}$ are the (anti)-fundamental quark superfields of the MSB sector,
$\tilde{f}$, $f$ and $\tilde{H}$, $H$
are the messengers and the Higgs fields transforming in the (anti)-fundamental of the $SU(5)$ GUT.
The factors of $1/16 \pi^2 $ on the right hand side of \eqref{supone} indicate
that these contributions come from loop effects in the underlying theory at the scale $M_p$. 
The constants $a_{i}$ are undetermined in the low-energy effective theory, they are generically of order one 
(which we will interpret as being in the range $10^{-3}\div 10^1$). 
These are the leading-order higher-dimensional operators which involve interactions between $WW$ and the
matter-field bilinear gauge singlets.
Operators of even higher dimension will be suppressed by extra powers of the Planck mass $M_p$ and will not
be relevant for our analysis.

The R-sector is described by a non-Abelian gauge theory. We will take it to be an SQCD theory
with the gauge group $SU(N_c)$ and $N_f$ flavours of quarks $\tilde{Q}_{\sst R}$, ${Q}_{\sst R}$
with $N_f < N_c-1$. The quark fields $\tilde{Q}_{\sst R}$, ${Q}_{\sst R}$ develop (large) VEVs
which break the gauge group to
$SU(N_c-N_f)$. The resulting `low-energy' theory of the R-sector is the pure $SU(N_c-N_f)$ SYM with the
dynamical scale $\Lambda_{\sst R}$ (plus colour-singlet meson fields $\tilde{Q}_{\sst R}{Q}_{\sst R}$).
The SYM theory is strongly-coupled at the scale $\Lambda_{\sst R}$ and develops a gaugino
condensate,
\be
\langle W_R^2 \rangle \, =\, \langle \lambda_R^2 \rangle \, =\, \Lambda_{\sst R}^3 \ .
\label{g-cond}
\ee
This effect in the superpotential \eqref{supone} generates masses $m_{Q_{MSB}}$, $m_f$ and  $m_{H}$
of the order $\sim{\Lambda_{\sst R}^3}/{M_p^2}$
for the appropriate chiral matter fields.
This mass generation is the retrofitting mechanism of \cite{DM} as explored recently in \cite{AS}
in the ISS model building context. A novel feature of our model compared to \cite{AS}
is the fact that in our context not only the MSB-quarks and the messengers, 
but also the GUT Higgs fields $H$ and $\tilde{H}$ get a retrofitted 
mass $m_{H}$ which
gives rise to the $\mu_{\sst MSSM}$ parameter of the Standard Model,
\be
\mu_{\sst MSSM} \,\equiv \, m_{H} = \frac{a_{3}}{16\pi^2} \frac{\Lambda_{\sst R}^3}{M_p^2} \ .
\label{retrofit}
\ee

The generation of the quark masses $m_{Q_{MSB}}\sim{\Lambda_{\sst R}^3}/{M_p^2}$ 
is a key ingredient for the metastable susy breaking \cite{ISS}
in the MSB sector. The relevant scale is \cite{AS}:
\be
\mu^2_{\sst MSB} \,\equiv \, \Lambda_{\sst MSB} m_{Q_{MSB}} = 
\frac{a_{1}}{16\pi^2} \frac{\Lambda_{\sst MSB}\Lambda_{\sst R}^3}{M_p^2} \ .
\label{mmsbret}
\ee
In the context of our model, the generation of $\mu_{\sst MSSM}$ in \eqref{retrofit} 
and $\mu_{\sst MSB}$ in \eqref{mmsbret}
are the only relevant effects of the 
retrofitted superpotential \eqref{supone}.  The value of $\Lambda_{\sst R}\ \gtrsim\, 10^{14}$ GeV 
is then chosen, so as to give
\be
\mu_{\sst MSSM} =\frac{a_{3}}{16\pi^2}  \frac{\Lambda_{\sst R}^3}{M_p^2} \, \gtrsim \, 10^2 
\, \div \, 10^3  \, {\rm GeV} ,\ .
\label{muretr}
\ee
as required for electro-weak symmetry breaking. 

Although the messenger fields $f,\tilde{f}$ also get a contribution to their masses 
from ${\cal W}_1$, the dominant contribution to $m_f$  comes from a second 
class of interactions between gauge singlets from different sectors.
These couple the messenger fields of the GUT 
sector and the quark bilinears from the hidden sectors;
\be
{\cal W}_2 \, =\,
\frac{ b_{1}}{M_p}\,\tr(\tilde{f}f)\, \tr(\tilde{Q}_{\sst MSB} Q_{\sst MSB})
\,+ \, \frac{b_{2}}{M_p}\, (\tilde{f}f) \, (\tilde{Q}_{\sst R} Q_{\sst R})
\ ,
\label{suptwo}
\ee
with constants $b_{1},\,b_{2}$.
These terms are ultimately responsible for the mediation of susy-breaking from the MSB-sector
to the GUT-sector, and specifically for the generation of Majorana gaugino masses.
The traces in \eqref{suptwo} are over gauge and flavour indices of each sector. Furthermore,
as mentioned earlier, the flavour symmetry $SU(N_f=5)$ of the R-sector is gauged, and this
makes the R-meson field $\tilde{Q}_{\sst R} Q_{\sst R}$ an adjoint plus a singlet
under the GUT $SU(5)$ gauge group,
\be
\Phi_{\sst GUT}^{ij} \, =\, 
\frac{1}{\langle \tilde{Q}_{\sst R} Q_{\sst R}\rangle^{\frac{1}{2}}}\,
 \tilde{Q}_{\sst R}^i Q_{\sst R}^j   \ , 
\qquad i,j = 1 \ldots N_f=5\ .
\label{def-PGUT}
\ee 
We will show in the next subsection that the VEV for $\tilde{Q}_{\sst R}^i Q_{\sst R}^j$ is
generated dynamically in the R-sector of our theory and is of the form
\be
 \tilde{Q}_{\sst R}^i Q_{\sst R}^j   \, =\, M_{\sst GUT}^2 \, {\rm diag}(+1,+1,+1,-1,-1) \ , 
 \ .
 \label{vevvev}
\ee
The mass term for the messengers arises from the last term\footnote{The structure 
of the last term in \eqref{suptwo} is a short-hand for a generic interaction,
consistent with an unbroken $SU(5)$,
$f \cdot(c_1 \tr(\tilde{Q}_{\sst R} Q_{\sst R})+c_2\tilde{Q}_{\sst R} Q_{\sst R})\cdot \tilde{f}$,
where $c_{1,2}$ are constants of order 1.}
 in \eqref{suptwo}. Using \eqref{vevvev} we find
\be
m_f = b_{2}\frac{M^{2}_{\sst GUT}}{M_p}\,.
\label{mmess}
\ee
The third class of interactions couples the Higgs (anti)-fundamental fields of the GUT sector
to the adjoint (plus a singlet) Higgs which arises from mesons of the R-sector. It has the form,
\be
{\cal W}_3 \, =\, 
\frac{\kappa}{M_p}\, H \cdot \left(
\tr(\tilde{Q}_{\sst R} Q_{\sst R}) \, + \, \tilde{Q}_{\sst R} Q_{\sst R}\right) \cdot \tilde{H} \  .
\label{supthree}
\ee
These two terms are included to raise the mass of the Higgs triplet fields and do not give 
any additional mass to the doublets. In order for this to be the case we require
the couplings to be precisely equal as shown. The doublet-triplet splitting will be discussed in 
more detail below.

\subsection{R-sector and the generation of the GUT scale} \label{sec:Rsect}

In our approach all mass-parameters
should be generated dynamically.
An important point then is to explain how the GUT scale $M_{\sst GUT} \sim 10^{16} \div 10^{17}$ GeV
is generated alongside the much lower $\mu_{\sst MSSM}$ scale in \eqref{muretr}.  
In this sub-section we will show that this hierarchy of scales is
naturally explained by the dynamics of the R-sector of our model.

As already mentioned, the R-sector is given by an SQCD with $N_c > N_f +1$,
with the number of flavours being set to $N_f=5.$
The quarks are exactly massless since in the general set-up which we follow no tree-level
masses can be put in by hand. As is well-known, there is a nonperturbative 
Affleck-Dine-Seiberg superpotential \cite{ADS} in this theory which leads to run-away
vacua and renders the theory inconsistent, unless there is a mechanism to prevent the
run-away and stabilize the vacua. Without loss of generality and naturalness, this
is easily  achieved by adding a
leading-order higher-dimensional operator to the Lagrangian,
\be
\frac{d}{2 M_p} \tr(\tilde{Q}_{\sst R} Q_{\sst R})^2 \ ,
\ee
where $d$ is a constant,
so that the total
superpotential for the meson fields of the R-sector is,
\be
{\cal W}_{\sst R} \, =\, (N_c-N_f)\left(
\frac{\Lambda_{\sst SQCD}^{3N_c-N_f}}{\det_{N_f}(\tilde{Q}_{\sst R} Q_{\sst R})}
\right)^{\frac{1}{N_c-N_f}} \, + \, \frac{d}{2 M_p} \tr(\tilde{Q}_{\sst R} Q_{\sst R})^2 \ .
\label{supR}
\ee
The dynamical scale $\Lambda_{\sst SQCD}$ appearing in the Affleck-Dine-Seiberg superpotential
above, is the scale of the full SQCD theory of the R-sector, and should be distinguished from
the dynamical scale $\Lambda_{\sst R}$ of the `low-energy' $SU(N_c-N_f)$ pure SYM. The relation
between $\Lambda_{\sst SQCD}$ and $\Lambda_{\sst R}$ will be determined below.

In terms of the meson field $M_{ij} =\tilde{Q}_{\sst R}^i Q_{\sst R}^j$
the F-flatness condition on \eqref{supR} gives an equation for diagonal components
(without loss of generality we work in the basis where $\langle M_{ij} \rangle$ is diagonal),
\be
\label{mscale}
\langle M_{ii}  \rangle^2 \, =\, 
\frac{M_{p}}{d} \left(\frac{\Lambda^{3N_{c}-N_{f}}_{\sst SQCD}}{\det_{N_f}\, M}\right)^\frac{1}{N_c-N_f} \ ,
\ee
which holds for each value of $i=1,\ldots,N_f=5$.
Since the right hand side of \eqref{mscale} does not
depend on $i$ it follows that all the values of $\langle M_{ii} \rangle^2$ 
must be equal to each other. However this
does not necessarily imply that the VEVs of the meson field itself are all the same. 
For $N_f=5$ there are three inequivalent discrete solutions of \eqref{mscale},
the first one is
\be
\langle M_{ij} \rangle \, =\, \langle M \rangle \, {\rm diag} (+1,+1,+1,+1,+1) \quad => \quad SU(5) \ ,
\ee
the second solution breaks $SU(5)$ down to $SU(4)$,
\be
\langle M_{ij} \rangle \, =\, \langle M \rangle \, {\rm diag} (+1,-1,-1,-1,-1) \quad => \quad   SU(4) \ ,
\ee
while the third solution is precisely what we require, 
it corresponds to a spontaneous breakdown of $SU(5) \to SU(3) \times SU(2) \times U(1)$,
\be
\langle M_{ij} \rangle \, =\, \langle M \rangle \, {\rm diag} (+1,+1,+1,-1,-1) 
\quad => \quad  SU(3) \times SU(2) \times U(1) \ .
\ee

The vacuum expectation value of the meson field in \eqref{mscale} should now be expressed 
in terms of the dynamical scale $\Lambda_{\sst R}$ for the effective pure SYM 
$SU(N_c-N_f)$ theory. This is easily achieved by 
matching the gauge couplings of the SQCD and the SYM theories at the scale $\sqrt{M}$,
\begin{equation}
 \Lambda^{3N_{c}-N_{f}}_{\sst SQCD}=\Lambda^{3(N_{c}-N_{f})}_{\sst R}M^{N_{f}}.
\end{equation}
Inserting into Eq. \eqref{mscale} gives
\be
\langle M \rangle =\frac{1}{\sqrt{d}} \sqrt{\Lambda_{\sst R}^3 M_p} \ ,
\ee
in terms of $\Lambda_{\sst R}$, which is precisely what we are after.

Finally, we need to define a canonically normalised meson field $\Phi_{\sst GUT}$
in terms of the dimension-two meson field we were using so far. There are essentially two dimensionful
parameters, $\sqrt{\langle M \rangle}$ and $\Lambda_{\sst R}$ in the QCD theory of the R-sector,
which obey $\sqrt{\langle M \rangle} \gg \Lambda_{\sst R}.$
The first parameter sets the scale where the full $SU(N_c)$ is broken down to $SU(N_c-N_f)$, and the
second parameter, is the confinement scale of the $SU(N_c-N_f)$ SYM.
The mesons describe the Higgsing of $N_{f}$ of the $N_{c}$ colors. Under the remaining $SU(N_{c}-N_{f})$
they are colour neutral and they do not take part in the confinement of the $SU(N_{c}-N_{f})$.
Hence, the appropriate scale is the scale $\sqrt{\langle M \rangle}$ at which the gauge group is Higgsed 
\footnote{In the first version of this paper we argued that we can expand the Kahler 
potential for the mesons as 
\mbox{$K\sim\frac{\rm{const} }{\Lambda_{R}}M^{\dagger}M+\frac{\rm const}{\sqrt{\langle M\rangle}}M^{\dagger}M$}. 
If the constant in the first term is non-zero the first term dominates and we would have to 
normalise with $\Lambda_{\sst R}$. However, since the mesons do not couple to
the remaining $SU(N_{c}-N_{f})$ the first term vanishes. Another way to see that the normalisation
\eqref{def-PGUT2} is the right one is to note that the masses of $SU(5)$ vector bosons 
$m_{v} \sim g \langle Q_{\sst R} \rangle \sim g \langle M \rangle^{\frac{1}{2}}$
should be the same whether we think of the $SU(5)$ being higgsed by quarks $Q_{\sst R}$ or by 
mesons $\Phi_{\sst GUT}$. } is
\be
\Phi_{\sst GUT}^{ij} \, =\, \frac{1}{\sqrt{\langle M\rangle} }\,
\tilde{Q}_{\sst R}^i Q_{\sst R}^j  \ .
\label{def-PGUT2}
\ee 

In total we have
\be
\langle \Phi_{\sst GUT}^{ij} \rangle \, =\, M_{\sst GUT} \, {\rm diag}(+1,+1,+1,-1,-1) \ , 
\ee
where
\be
M^{2}_{\sst GUT}\sim \langle M\rangle = \frac{1}{\sqrt{d}}\sqrt{\Lambda^{3}_{\sst R} M_p} \ .
\label{mgut2}
\ee
Eliminating $\Lambda_{R}$ with \eqref{muretr} we arrive at a relation 
between $\mu_{\sst MSSM}$ and $M_{\sst GUT}$ as anticipated in the 
Introduction;
\be
M^{2}_{\sst GUT}=\frac{4\pi}{\sqrt{a_{3}\,d}}\left(\mu_{\sst MSSM} M^{3}_{p}\right)^{1/2} \ .
\label{gutmurel-2}
\ee
Taking $M_{p}\sim 10^{19}\,\rm{GeV}$ and $\mu_{\sst MSSM}\sim 10^{2}\div 10^{3}\, \rm{GeV}$ we find
\be
M_{\sst GUT}\, \sim\,  10^{15} \div 10^{17}\, {\rm GeV} \ ,
\ee
if we choose the constants $a_{3},\,d$ in the range $10^{-3}\div 10^1$.

\subsection{Metastable supersymmetry breaking} \label{sec:msb}

The MSB sector is described by the ISS \cite{ISS} model which is an SQCD with 
$N_f$ flavours of classically massless quarks and $N_c+1 \le N_f < 3N_c/2$. The quarks 
$\tilde{Q}_{\sst MSB}$, ${Q}_{\sst MSB}$ generate masses 
dynamically via the interactions
\eqref{supone}
with the R-sector as explained above. 

Following ISS \cite{ISS} we introduce canonically normalised fields 
\begin{equation}
\label{normalised}
\Phi_{\sst MSB}\,=\,
\frac{\tilde{Q}_{\sst MSB}Q_{\sst MSB}}{\Lambda_{\sst MSB}} \ .
\end{equation}
The magnetic description of the gauge theory, then has a classical
\be
{\mathcal{W}}_{\rm cl}\, =\, h\, \tr_{\sst \Nf} \p \Phi_{\sst MSB} \pt\, -\, 
h\mu^{2}_{\sst MSB}\, \tr_{\sst \Nf} \Phi_{\sst MSB} \ ,
\label{Wcl}
\ee
and dynamical superpotential
\be
{\mathcal{W}}_{\rm dyn}\, =\, 
N\left( h^\Nf \frac{\det_{\sst \Nf} \Phi_{\sst MSB}}{\Lambda_{\sst MSB}^{\Nf-3N}}\right)^\frac{1}{N}\ ,
\label{Wdyn}
\ee
where $N=\Nf-\Nc$ and $h$ is a constant. Moreover, $\pt$ and $\p$ are the magnetic quarks made up from 
suitable combinations of $\tilde{Q}_{\sst MSB}$ and $Q_{\sst MSB}$.
Using the normalisation \eqref{normalised} one easily translates the retrofitted mass term for 
$\tilde{Q}_{\sst MSB}$ and $Q_{\sst MSB}$,
$m_{Q_{\sst MSB}}\sim \Lambda_{\sst R}^{3}/(16 \pi^{2} M^{2}_{p})$ 
into $\mu^{2}_{\sst MSB}$ as given in Eq. \eqref{mmsbret}.

In the metastable vacuum near $\Phi_{\sst MSB}=0$ supersymmetry is broken by the rank condition at 
the scale $\mu_{\sst MSB}$. In particular, we have
\begin{equation}
 \tr (F_{\Phi^{ij}_{\sst MSB}})\, \sim \,\mu^2_{\sst MSB} \ .
\end{equation}

This supersymmetry breaking is then gauge mediated to the GUT sector 
by the messengers $\tilde{f},f$ and the interaction to $\Phi_{\sst MSB}$ arising from the first part of Eq. \eqref{suptwo}. 
The usual one-loop diagram with messengers propagating in the loop,
generates Majorana mass terms for the gauginos of the GUT-sector,
\begin{equation}
 m_{\lambda}\, \sim\,  b_{1} \frac{g^{2}}{16\pi^{2}}\frac{\Lambda_{\sst MSB}}{M_{p}}\frac{\tr(F_{\Phi_{\sst MSB}})}{m_{f}}
 \, \sim\, \frac{g^{2}}{16 \pi^{2}}\frac{a_{1}b_{1}}{a_{3}b_{2}}
 \left(\frac{\Lambda_{\sst MSB}}{M_{\sst GUT}}\right)^{2} \mu_{\sst MSSM}\ . 
 \label{gaumass}
\end{equation}
In the above equation $\Lambda_{\sst MSB}$ is a free parameter, and it can always be set such that
the values of the gaugino masses are in the desired range,
\begin{equation}
 m_{\lambda}\sim 1\,\rm{TeV} \ .
\end{equation}

Stability of the MSB sector requires that the messengers are non-tachyonic \cite{MN},
\begin{equation} 
 b_{1}\frac{\Lambda_{\sst MSB}}{M_{p}}\mu^{2}_{\sst MSB}
 \,<\,  m^{2}_{f}=\left(b_{2}\frac{M^{2}_{\sst GUT}}{M_{p}}\right)^{2}\ ,
\end{equation}
and that tunneling to a possible supersymmeric vacuum with $\langle \tilde{f}\rangle,\langle f\rangle\neq 0$ is slow,
\begin{equation}
 \frac{b_{2}}{b_{1}}\frac{M^{2}_{\sst GUT}}{\Lambda_{\sst MSB}}\, \gg\,  \mu_{\sst MSB}\ .
\end{equation}
Both conditions can be fulfilled in our model.
Similarly, possible flavor changing effects caused by gravity mediation can be made small for a suitable choice of constants,
\begin{equation}
 m_{3/2}\, =\, \frac{\mu^{2}_{\sst MSB}}{M_{p}}\lesssim 10^{-2} m_{\lambda}\ .
\end{equation}

\subsection{Doublet-triplet splitting and proton decay }

Let us return to the Higgs sector and in
particular the Higgs triplets. First we should mention that the main
issue with minimal $SU(5)$ GUTs is that they can predict too rapid proton
decay because of dimension-5 operators generated by terms of the form
$QQQL$ or $U^{c}U^{c}D^{c}E^{c}$ in the effective tree-level superpotential
(\cite{dim51,dim52}, see \cite{Raby:2006sk,Nath:2006ut} for a review). This question
is also important for our model as we shall now see.

The Higgs triplets are made heavy by the effective operator 
\be
{\cal W}_{3}\, =\, \kappa\, \frac{M_{\sst GUT}}{M_{p}}\, 
H\cdot (tr(\Phi_{\sst GUT})+\Phi_{\sst GUT})\cdot\tilde{H} \ ,
\ee
where $\kappa$ represents an unknown constant, and their masses 
are therefore of order
\be
m_{H_{3},\bar{H}_3}\approx \kappa M_{\sst GUT}^2/M_{p}.\ee
Note that the effective mass is proportional to 
$\tr(\Phi_{\sst GUT})+\Phi_{\sst GUT}=2\,\mbox{diag}(1,\,1,\,1,\,0,\,0),$
so that the combined coupling shares some features with the Dimopoulos-Wilczek
form as discussed widely in the context of $SO(10)$ 
\cite{dimwil,Babu:1993we,Babu:2002fs,Kitano:2001ie,Kitano:2006wm}. Indeed our model
is rather more natural than standard minimal $SU(5)$ for 
precisely the same reason as $SO(10)$, namely because the 
meson field $M$ is not traceless. We would also argue that the requirement 
that the two couplings in ${\cal W}_3$ be identical could conceivably be met  
by the underlying physics and is a less distasteful fine-tuning than that which 
occurs in minimal $SU(5)$. (Note that the renormalization of both couplings is 
identical in the fully supersymmetric theory.)

In order to see if one can avoid dimension-5 operators 
that are too large, we must consider how the 
spectrum influences the possible values of $M_{\sst GUT}$.
For reference we now collect the relevant mass-scales. First as 
we have seen $M_{\sst GUT}$ itself is set by $\Lambda_{\sst R}$ and 
subsequently $\mu_{\sst MSSM}$ to lie in the range 
$10^{15.5}\lesssim M_{\sst GUT}\lesssim 10^{17}$GeV. 
The mass spectrum of the Higgs sector of in model is as follows.
The fundamental Higgses $H$ and $\tilde{H}$ split into a doublet and a triplet parts
\begin{eqnarray}
m_{H_2,\, \bar{H}_2} & \sim & \mu_{\sst MSSM} \ , \\
m_{H_3,\, \bar{H}_3} & \approx & \kappa \frac{M_{\sst GUT}^2}{M_p} \ .
\end{eqnarray}
At and below the GUT scale the light degrees of freedom contained in the elementary 
quarks $Q_{\sst R}$ and $\tilde{Q}_{\sst R}$ of the R-sector are naturally packaged into the
composite R-meson Higgs $\Phi_{\sst GUT}$. It contains the unrealised Goldstone bosons (eaten by the
massive GUT vector bosons) as well as the weak-triplet fields $\sigma_3$, colour-octet
fields $\sigma_8$ and the singlet. In our model the masses $m_{\sigma_3}$, $m_{\sigma_8}$
and $m_{{\bf 1}} $ are the same and given by
\be
m_{\sigma_3,\, \sigma_8,\, {\bf 1}} 
\approx  \frac{\Lambda_{\sst R}^3}{M_{\sst GUT}^2} 
\approx 2 d \frac{M_{\sst GUT}^2}{M_p} \, . \label{trip-oct}
\ee
Here $H_2$ and $H_3$ denote the doublet and the triplet parts of the fundamental Higgs $H$
of the $SU(5).$ 
One requires $m_{H_3,\, \bar{H}_3} \gtrsim 7\times 10^{16}$GeV
in order to avoid proton decay \cite{Murayama:2001ur}. 
This can be achieved with a 
moderately large value of $\kappa\sim 10$ and a GUT scale 
at the high end of the range,
but the weak-triplets and colour-octets get only 
$F$-term masses which are significantly less than 
$M_{\sst GUT}$. Indeed, recall that for the range of $M_{\sst GUT}$ 
that we are considering, 
we have $10^{-6}\lesssim a_3 d \lesssim 1$ as determined 
by $\mu_{\sst MSSM}$, but $d$ itself can be kept as an 
essentially free parameter. 

As discussed in 
\cite{Bachas:1995yt,Chkareuli:1998wi,Bajc:2002bv,Bajc:2002pg,Bajc:2006pa}
unification at values of $M_{\sst GUT}$ that are greater than the 
canonical value of $10^{16}$~GeV are possible if 
$ m_{\sigma_3,\, \sigma_8,\, {\bf 1}} \ll M_{\sst GUT} $ which is 
generically true for this model.
A general analysis of the gauge-coupling RGE's in $SU(5)$  yields 
two relations that we should satisfy in order to preserve unification
\cite{Bajc:2002bv,Bajc:2002pg,Bajc:2006pa};
\begin{eqnarray}
M_{\sst GUT} & = & M_{\sst GUT}^0 \left( \frac{M_{\sst GUT}^0}{
 m_{\sigma_3,\, \sigma_8} } \right)^{\frac{1}{2}} \ ,
 \label{first-c} \\
 m_{H_3,\, \bar{H}_3} & = & m_{H_3,\, \bar{H}_3}^0 \left( \frac{m_{\sigma_3}}{m_{\sigma_8}}\right)^{\frac{5}{2}}\, ,
\label{second-c}
\end{eqnarray}
where $M^0_{\sst GUT}=m_{H_3,\, \bar{H}_3}^0=10^{16} $ GeV are the values 
at the usual unification scale when one assumes a desert between $M_{\sst SUSY}$ and $M_{\sst GUT}$. In terms of 
the coupling $d$ (which appears in \eqref{trip-oct}) the first requirement \eqref{first-c}
is rather interesting: it becomes
\be
M_{\sst GUT} \approx \, \frac{(M_{GUT}^{0})^{\frac{3}{4}} M_p^{\frac{1}{4}}}{d^{\frac{1}{4}}}\, .
\ee
It is a remarkable fact that in our model the 
determination of $\mu_{\sst MSSM}$ fixes
$M_{\sst GUT}\propto ({a_3 d})^{-\frac{1}{4}}$ as 
we have already noted in \eqref{gutmurel-2}. Thus as long as we set $a_3$ so that we 
get $ m_{\sigma_3,\, \sigma_8} =M_{\sst GUT}^0$ at the usual unification 
scale of $10^{16}$~GeV, we may treat $d$ as an independent parameter which splits 
$M_{\sst GUT}$ and $ m_{\sigma_3,\, \sigma_8} $ in the right way. 
The required value of $a_3$ can be taken to be
$a_3d\sim 1$ for "usual" unification at $M_{\sst GUT}^0$ (note that the 
values are extremely sensitive to adjustments in $M_{\sst GUT}^0$ so 
the discussion at this point is very qualitative), 
and since $ m_{\sigma_3,\, \sigma_8} =M_{\sst GUT}^0$ 
for this unification, we have $d=M_p/M_{\sst GUT}^0\approx 10^3 $ 
and hence $a_3\approx 10^{-3}$. We can then 
scale $d$ independently and the first relation \eqref{first-c} is always satisfied. 
In particular $d=10^{-3}$ then gives $M_{\sst GUT}\sim 10^{17}$GeV and 
$m_{\sigma_3,\, \sigma_8} \sim 10^{12}$~GeV.

Unfortunately in the model presented here we have
$ m_{\sigma_3} = m_{\sigma_8} $ so the second relation \eqref{second-c} requires 
$m_{H_3,\, \bar{H}_3}^0=10^{16} $ GeV. In other words making the 
triplets heavy enough to avoid proton decay by adjusting 
$\kappa $ is incompatible with exact gauge unification for this model. 
It is unclear whether more precise study of this issue including 
for example two-loop effects would change this conclusion. 

In addition, of course, if one is willing to go to product
{}``GUT'' groups, such as Pati-Salam models, or models based on
flipped $SU(5)$, then the doublet-triplet mass splitting problem
can be easily avoided. In the latter case for example, the GUT symmetry
is broken by VEVs of a $\mathbf{10}$ and $\mathbf{\bar{10}}$ 
rather than an adjoint Higgs, and the doublet and triplet masses are
automatically split. Unfortunately in this case one would have to
abandon the adjoint of $SU(N_{f})$ which arose rather nicely from
the confinement of $SU(N_{c})$. Also it is unclear how $\mathbf{10}$'s
and $\mathbf{\bar{10}}$'s would appear as composite fields in the  superpotential
of the $R$-sector.

Given the similarity of the coupling to the Dimopoulos-Wilczek solution 
to the doublet-triplet problem, a natural avenue to explore \cite{WIP} in this class of 
models is embedding the $SU(5)$ structure within $SO(10)$.
In fact all the main results of this paper can be straightforwardly generalised to 
an $SO(10)$ Grand Unified Theory, and are not specific to the minimal $SU(5)$.
 The reader is referred to \cite{Raby:2006sk} for further references 
to the doublet-triplet mass-splitting problem.

\section{Discussion} \label{sec:disc}

We have presented an extremely compact formulation of a supersymmetric Grand Unified $SU(5)$ theory.
Our model has the following features:

Supersymmetry is broken spontaneously by a long-lived metastable vacuum state of a hidden MSB sector.
This supersymmetry breaking is communicated to the GUT theory via gauge mediation and generates
gaugino masses which can be made $\sim 10^2 \div 10^3$ GeV. Squark, slepton and higgsino mass splittings follow from 
this in the standard gauge mediation way.

The model is fully natural with all mass-parameters generated dynamically via the retrofitted couplings
to the gluino condensate of the R-sector. In particular, by choosing the dynamical scale of the R-sector to be
$\Lambda_{\sst R} \sim 10^{14}$ GeV, we generate the $\mu$-parameter of the Standard Model,
$\mu_{\sst MSSM} \sim 10^2 \div 10^3$ GeV, which in turn generates the required electro-weak symmetry breaking scale
$\sim 10^2$ GeV.

Remarkably, the GUT scale $M_{\sst GUT} \sim 10^{15} \div 10^{17}$ GeV $\gg \mu_{\sst MSSM}$ is also dynamically generated
in our model. This follows from the fact that
the adjoint Higgs required in the GUT sector is identified with the traceless part of the meson matrix of the R-sector.
The GUT $SU(5)$ group arises from gauging the $SU(5)$ flavour group of the R-sector, and we show that
the required spontaneous breaking of $SU(5) \to SU(3) \times SU(2) \times U(1)$ does occur at the
scale $M_{\sst GUT} \sim (\Lambda_{\sst R}^3 M_p/d)^{\frac{1}{4}} \sim 10^{15} \div 10^{17}$ GeV.

Hence we have presented a simple and natural (modulo proton decay) model of susy GUT
which can explain the values of the symmetry-breaking scales and their hierarchies.
The model is weakly coupled and fully calculable including the soft-susy breaking terms.

\section*{Acknowledgements}   

 We thank Sakis Dedes and Stefan F\"orste for useful discussions. We are also grateful to 
 Borut Bajc, Pavel Perez and Goran Senjanovic for discussions and comments on an earlier version 
 of this paper.

\begin{appendix}
\section{Bounds on additional Planck suppressed operators}
In Sect. \ref{sec:model} we have made a certain selection among the Planck suppressed operators. In this appendix we will look at the interactions up to $1/M^{2}_{p}$ that we have neglected so far.

Let us start with the operators that involve gauge fields as well as matter fields as in \eqref{supone}. 
In Eq. \eqref{supone} we have neglected three types of operators,
\begin{eqnarray}
 \Delta{\mathcal{W}}_{1}&=&\frac{\lambda_{1}}{16\pi^{2} M^{2}_{p}} \left[
\tr\left(W^{2}_{\sst R}\right)\tr(\tilde{Q}_{\rm{R}}Q_{\rm{R}}) \right],
\\ 
\Delta{\mathcal{W}}_{2}&=&\frac{\lambda_{2}}{16\pi^{2} M^{2}_{p}} \left[
\tr\left(W^{2}_{\sst MSB}\right)\tr(\tilde{X}X) \right],
\\
 \Delta{\mathcal{W}}_{3}&=&\frac{\lambda_{3}}{16\pi^{2} M^{2}_{p}} \left[
\tr\left(W^{2}_{\sst GUT}\right)\tr(\tilde{X}X) \right],
\end{eqnarray}
where the $X$ is symbolic for all possible matter fields $Q_{\rm{R}},Q_{\sst MSB},f,H$.
$\Delta {\mathcal{W}}_{2}$ and $\Delta {\mathcal{W}}_{3}$ are harmless because $\tr(W^{2}_{\sst MSB})$
and $\tr(W^{2}_{\sst GUT})$ do not acquire significant vacuum expectation values.
$\Delta {\mathcal{W}}_{1}$ gives a mass of the order of 
$\Lambda^{3}_{\rm{R}}/(16\pi^{2}M^{2}_{p}) \sim \mu_{\rm{MSSM}}$ to the $Q_{\rm{R}}$ fields.
However, this term has to be compared to the second term of \eqref{supR} which also appears in the F-term for the GUT-field. Inserting the vacuum expectation value for 
$\tilde{Q}_{\rm{R}}Q_{\rm{R}}\sim M^{2}_{\sst GUT}$ we find that $\Delta{\mathcal{W}}_{1}$  is suppressed by a factor of $(M_{\sst GUT}/(4\pi\,M_{p}))^{2})\lesssim 10^{-5}$ compared to 
${\mathcal{W}}_{\rm{R}}$.
Therefore $\Delta{\mathcal{W}}_{1}$ is harmless as well. Overall,
\begin{equation}
 \lambda_{1},\lambda_{2},\lambda_{3}\,\, \rm{can}\,\,\rm{be}\,\,{\mathcal{O}}(1).
\end{equation}

The second class of possible additional operators involves four matter fields as in \eqref{suptwo} or \eqref{supthree}
and is suppressed by one power of $1/M_{p}$.
We have the following possibilities,
\begin{eqnarray}
\Delta {\mathcal{W}}_{4}&=&\frac{\lambda_{4}}{M_{p}} \left[\tr(\tilde{Q}_{\rm{R}}Q_{\rm{R}})\,\tr(\tilde{Q}_{\sst MSB}Q_{\sst MSB})\right],
\\
\Delta {\mathcal{W}}_{5}&=&\frac{\lambda_{5}}{M_{p}} 
\left[\tr\left([\tilde{Q}_{\sst MSB}Q_{\sst MSB}]^{2}\right)
+c_{5}\left[\tr(\tilde{Q}_{\sst MSB}Q_{\sst MSB})\right]^{2}\right],
\\
\Delta {\mathcal{W}}_{6}&=&\frac{\lambda_{6}}{M_{p}}
\left[\tr(\tilde{Q}_{\sst MSB}Q_{\sst MSB})\,\tr(\tilde{H}H)\right],
\\
\Delta {\mathcal{W}}_{7}&=&\frac{\lambda_{7}}{M_{p}} 
\left[\tr\left([\tilde{H}H]^{2}\right)+c_{7}\left[\tr(\tilde{H}H)\right]^{2}\right],
\\\
\Delta {\mathcal{W}}_{8}&=&\frac{\lambda_{8}}{M_{p}} 
\left[\tr(\tilde{H}H\tilde{f}f)+c_{8}\,\tr(\tilde{H}H)\,\tr(\tilde{f}f)\right],
\\
\Delta {\mathcal{W}}_{9}&=&\frac{\lambda_{9}}{M_{p}} 
\left[\tr\left([\tilde{f}f]^{2}\right)+c_{9}\left[\tr(\tilde{f}f)\right]^{2}\right].
\end{eqnarray}

Inserting the VEV $\langle\tilde{Q}_{\rm{R}}Q_{\rm{R}}\rangle \sim \Lambda_{\rm{R}}M_{\sst GUT}$ we find that $\Delta{\mathcal{W}}_{4}$ gives an additional contribution,
\begin{equation}
 \Delta \mu^{2}_{\sst MSB}=\lambda_{4}\frac{M^{2}_{\sst GUT}}{M_{p}}\Lambda_{\sst MSB}
\gtrsim 10^{11} \lambda_{4}\mu^{2}_{\sst MSB}.
\end{equation}
To keep our MSB scale at the desired\footnote{One might consider the possibility that $\Delta {\mathcal{W}}_{4}$ gives indeed the dominant contribution to $\mu_{\sst MSB}$. However, it turns out that the Landau pole of the
MSB-sector, $\Lambda_{\sst MSB}$, is then too close to $\mu_{\sst MSB}$.} value we therefore have to require,
\begin{equation}
 \lambda_{4}\lesssim 10^{-11}.
\end{equation}

Interactions of the type $\Delta {\mathcal{W}}_{5}$ have two undesirable effects since they lead to linear
terms in the potential through 
$F_{\Phi_{\sst MSB}}=\mu^{2}_{\sst MSB}+\lambda_{5}(\Lambda^{2}_{\sst MSB}/M_{p})\,\Phi_{\sst MSB}+\ldots$.
This can either directly destabilise the metastable minimum or cause a shift in the messenger mass $M_{f}$ that, in turn
again destabilizes the SUSY breaking vacuum. This leads to the constraint \cite{MN},
\begin{equation}
\label{lambda5const}
 \frac{\lambda_{5}\Lambda^{2}_{\sst MSB}}{M_{p}}\lesssim
 \rm{min}\left[0.1\,\mu_{\sst MSB},
10^{-2}\frac{b_{1}M^{2}_{\sst GUT}M_{p}}{b_{2}\Lambda^{2}_{\sst MSB}}\right],
\end{equation}
 where $\lambda_{f}$ and $\tilde{\lambda}_{f}$ are the constants of order one in front of the first 
 and second term in Eq. \eqref{suptwo}. The first part of Eq. \eqref{lambda5const} is the more constraining and leads to
\begin{equation}
  \lambda_{5}\lesssim 10^{-2}.
\end{equation}

An interaction of type $\Delta {\mathcal{W}}_{6}$ would turn the Higgs fields into messengers. At first this looks like a very nice feature. Unfortunately, it also leads to a very large mass term for the Higgs field. This mass comes again from the contribution to the 
$F_{\Phi_{\sst MSB}}=\mu^{2}_{\sst MSB}+\lambda_{6}(M_{\sst GUT}/M_{p})\tilde{H}H+\ldots$. The cross terms lead to a contribution of  
\begin{equation}
 \Delta m^{2}_{H}=2\lambda_{6} \mu^{2}_{\sst MSB}\frac{M_{\sst GUT}}{M_{p}}.
\end{equation}
For the Higgs doublet that is part of $H$ and $\tilde{H}$ the mass must be of the order of the electroweak scale and we need
\begin{equation}
 \lambda_{6}\lesssim 10^{-15}.
\end{equation}

Neither $\tilde{H},H$ nor $\tilde{f},f$ aquire any significant (bigger than the electroweak scale) 
expectation values. Therefore the remaining interactions $\Delta{\mathcal{W}}_{7}$,$\Delta{\mathcal{W}}_{8}$ 
and $\Delta{\mathcal{W}}_{9}$ provide only additional Planck mass suppressed higher order interactions between the Higgses and the messengers. These interactions are not very constrained and
\begin{equation}
 \lambda_{7},\lambda_{8},\lambda_{9}\,\, \rm{can}\,\,\rm{be}\,\,{\mathcal{O}}(1).
\end{equation}

Overall the discussion of this appendix shows that the interactions
$\Delta{\mathcal{W}}_{4}$,$\Delta{\mathcal{W}}_{5}$ and $\Delta{\mathcal{W}}_{6}$ should be highly suppressed 
or, preferably, prevented by some mechanism of the underlying theory. 
All other terms can appear with their natural coefficients of order one.

\end{appendix}


\vspace{1cm}

\begin{thebibliography}{99}

\bibitem{ISS}
  K.~Intriligator, N.~Seiberg and D.~Shih,
  ``Dynamical SUSY breaking in meta-stable vacua,''
  JHEP {\bf 0604} (2006) 021
  hep-th/0602239.
  
  \bibitem{ACJK} 
S.~A.~Abel, C.~S.~Chu, J.~Jaeckel and V.~V.~Khoze,   
``SUSY breaking by a metastable ground state: Why the early universe preferred the non-supersymmetric vacuum,''   
JHEP {\bf 0701}, 089 (2007)   [arXiv:hep-th/0610334].

\bibitem{heat2} 
N.~J.~Craig, P.~J.~Fox and J.~G.~Wacker,   
``Reheating metastable O'Raifeartaigh models,''   arXiv:hep-th/0611006.

\bibitem{heat3} 
W.~Fischler, V.~Kaplunovsky, C.~Krishnan, L.~Mannelli and M.~Torres,   
``Meta-stable supersymmetry breaking in a cooling universe,''   arXiv:hep-th/0611018.

\bibitem{heat4} 
S.~A.~Abel, J.~Jaeckel and V.~V.~Khoze,   
``Why the early universe preferred the non-supersymmetric vacuum. II,''   
JHEP {\bf 0701}, 015 (2007)   [arXiv:hep-th/0611130]. 

\bibitem{heat5}
L.~Anguelova, R.~Ricci and S.~Thomas,   
``Metastable SUSY breaking and supergravity at finite temperature,''  
 arXiv:hep-th/0702168.   

\bibitem{Nelson:1993nf}
A.~E.~Nelson and N.~Seiberg,   
``R symmetry breaking versus supersymmetry breaking,''   Nucl.\ Phys.\  B {\bf 416}, 46 (1994)   
[arXiv:hep-ph/9309299].   

\bibitem{ELR}
  J.~R.~Ellis, C.~H.~Llewellyn Smith and G.~G.~Ross,
 ``Will The Universe Become Supersymmetric?,''
  Phys.\ Lett.\ B {\bf 114} (1982) 227.

\bibitem{dimop} 
S.~Dimopoulos, G.~R.~Dvali, R.~Rattazzi and G.~F.~Giudice,   
``Dynamical soft terms with unbroken supersymmetry,''   
Nucl.\ Phys.\  B {\bf 510} (1998) 12   [arXiv:hep-ph/9705307].   

\bibitem{Luty} M.~A.~Luty,   
``Simple gauge-mediated models with local minima,''   Phys.\ Lett.\  B {\bf 414} (1997) 71   [arXiv:hep-ph/9706554].   

\bibitem{Dine:2006gm} 
M.~Dine, J.~L.~Feng and E.~Silverstein,   
``Retrofitting O'Raifeartaigh models with dynamical scales,''   Phys.\ Rev.\  D {\bf 74}, 095012 (2006)   
[arXiv:hep-th/0608159].   

\bibitem{DM}
M.~Dine and J.~Mason,   
``Gauge mediation in metastable vacua,''   arXiv:hep-ph/0611312.   

\bibitem{Kitano:2006xg} 
R.~Kitano, H.~Ooguri and Y.~Ookouchi,   
``Direct mediation of meta-stable supersymmetry breaking,''   arXiv:hep-ph/0612139.   

\bibitem{Csaki:2006wi}
C.~Csaki, Y.~Shirman and J.~Terning,   
``A simple model of low-scale direct gauge mediation,''   arXiv:hep-ph/0612241.   


\bibitem{Abel:2007uq} 
S.~A.~Abel and V.~V.~Khoze,   
``Metastable SUSY breaking within the standard model,''   arXiv:hep-ph/0701069.   

\bibitem{MN}  
H.~Murayama and Y.~Nomura,   ``Gauge mediation simplified,''   arXiv:hep-ph/0612186.   

\bibitem{AS} 
O.~Aharony and N.~Seiberg,   ``Naturalized and simplified gauge mediation,''   
arXiv:hep-ph/0612308.   

\bibitem{Murayama:2007fe}
H.~Murayama and Y.~Nomura, ``Simple scheme for gauge mediation,''   arXiv:hep-ph/0701231.   


\bibitem{Nilles:1995ci}
H.~P.~Nilles,   ``Phenomenological aspects of supersymmetry,''   arXiv:hep-ph/9511313.   

\bibitem{Raby:2006sk}   S.~Raby,   
``Grand unified theories,''   arXiv:hep-ph/0608183.   

\bibitem{SGUT}
  S.~Dimopoulos and H.~Georgi,
  ``Softly Broken Supersymmetry And SU(5),''
  Nucl.\ Phys.\  B {\bf 193} (1981) 150. \\
  N.~Sakai,
  ``Naturalness In Supersymmetric Guts,''
  Z.\ Phys.\  C {\bf 11} (1981) 153.
  
\bibitem{ADS}
  I.~Affleck, M.~Dine and N.~Seiberg,
  ``Dynamical Supersymmetry Breaking In Supersymmetric QCD,''
  Nucl.\ Phys.\  B {\bf 241} (1984) 493.
 
\bibitem{dimwil}S.~Dimopoulos and F.~Wilczek,
  ``Incomplete Multiplets In Supersymmetric Unified Models,''
   NSF-ITP-82-07, unpublished; 
  ``Supersymmetric Unified Models,''
{\it  In *Erice 1981, Proceedings, The Unity Of The Fundamental Interactions*, 237-249}

\bibitem{Babu:1993we} 
K.~S.~Babu and S.~M.~Barr,
  ``Natural suppression of Higgsino mediated proton decay in supersymmetric SO(10),''
  Phys.\ Rev.\  D {\bf 48}, 5354 (1993)
  [arXiv:hep-ph/9306242].

\bibitem{Babu:2002fs}
K.~S.~Babu and S.~M.~Barr,
  ``Eliminating the d = 5 proton decay operators from SUSY GUTs,''
  Phys.\ Rev.\  D {\bf 65}, 095009 (2002)
  [arXiv:hep-ph/0201130].

\bibitem{Kitano:2001ie} 
R.~Kitano and N.~Okada,
  ``Dynamical doublet-triplet Higgs mass splitting,''
  Phys.\ Rev.\  D {\bf 64}, 055010 (2001)
  [arXiv:hep-ph/0105220].

\bibitem{Kitano:2006wm} 
R.~Kitano,
  ``Dynamical GUT breaking and mu-term driven supersymmetry breaking,''
  Phys.\ Rev.\  D {\bf 74}, 115002 (2006)
  [arXiv:hep-ph/0606129].

\bibitem{Goto:1998qg}  T.~Goto and T.~Nihei,   ``Effect of RRRR dimension 5 operator on the proton decay in the minimal   SU(5) SUGRA GUT model,''   Phys.\ Rev.\  D {\bf 59}, 115009 (1999)   [arXiv:hep-ph/9808255].   

\bibitem{Murayama:2001ur}H.~Murayama and A.~Pierce,   ``Not even decoupling can save minimal supersymmetric SU(5),''   Phys.\ Rev.\  D {\bf 65}, 055009 (2002)   [arXiv:hep-ph/0108104].   

\bibitem{dim51}
  S.~Weinberg,
  ``Supersymmetry At Ordinary Energies. 1. Masses And Conservation Laws,''
  Phys.\ Rev.\  D {\bf 26}, 287 (1982).

\bibitem{dim52}
  N.~Sakai and T.~Yanagida,
  ``Proton Decay In A Class Of Supersymmetric Grand Unified Models,''
  Nucl.\ Phys.\  B {\bf 197}, 533 (1982).

\bibitem{Nath:2006ut}
  P.~Nath and P.~F.~Perez,
  ``Proton stability in grand unified theories, in strings, and in branes,''
  arXiv:hep-ph/0601023.

\bibitem{Bachas:1995yt}
  C.~Bachas, C.~Fabre and T.~Yanagida,
  ``Natural gauge-coupling unification at the string scale,''
  Phys.\ Lett.\  B {\bf 370}, 49 (1996)
  [arXiv:hep-th/9510094].

\bibitem{Chkareuli:1998wi}
  J.~L.~Chkareuli and I.~G.~Gogoladze,
  ``Unification picture in minimal supersymmetric SU(5) model with string
  remnants,''
  Phys.\ Rev.\  D {\bf 58}, 055011 (1998)
  [arXiv:hep-ph/9803335].

\bibitem{Bajc:2002bv}
  B.~Bajc, P.~Fileviez Perez and G.~Senjanovic,
  ``Proton decay in minimal supersymmetric SU(5),''
  Phys.\ Rev.\  D {\bf 66}, 075005 (2002)
  [arXiv:hep-ph/0204311].

\bibitem{Bajc:2002pg}
  B.~Bajc, P.~Fileviez Perez and G.~Senjanovic,
  ``Minimal supersymmetric SU(5) theory and proton decay: Where do we stand?,''
  arXiv:hep-ph/0210374.


\bibitem{Bajc:2006pa}
  B.~Bajc and G.~Senjanovic,
  ``Proton decay, supersymmetry breaking and its mediation,''
  arXiv:hep-ph/0611308.

 \bibitem{WIP} 
S.~A.~Abel, S.~F\"orste, J.~Jaeckel and V.~V.~Khoze,   
{\it in preparation}.



\end{thebibliography}
\end{document}